\documentclass[12pt]{iopart}

\usepackage{graphicx}

\usepackage{color}
\usepackage{hyperref}
\hypersetup{colorlinks, linkcolor=blue}

\usepackage{iopams}  
\begin{document}

\title[Resonant radiation of oscillating two-color soliton molecues]{Resonant \emph{Kushi}-comb-like multi-frequency radiation of oscillating two-color soliton molecules}

\author{O Melchert$^{1,2}$, S Willms$^{1,2}$, I Oreshnikov$^{3}$, A Yulin$^4$, U Morgner$^{1,2}$, I Babushkin$^{1,2}$ and A Demircan$^{1,2}$}

\address{$^1$ Leibniz Universit\"at Hannover, Institute of Quantum Optics, Welfengarten 1, 30167 Hannover, Germany}
\address{$^2$ Leibniz Universit\"at Hannover, Cluster of Excellence PhoenixD, Welfengarten 1A, 30167 Hannover, Germany}
\address{$^3$ Max Planck Institute for Intelligent Systems, Max-Planck-Ring 4, 72076 T\"ubingen, Germany}
\address{$^4$ Department of Nanophotonics and Metamaterials, ITMO University, Birzhevaya line 16, 199034 St.\ Petersburg, Russia}


\begin{abstract}
Nonlinear waveguides with two distinct domains of anomalous dispersion can support the formation of molecule-like two-color pulse compounds. They consist of two tightly bound subpulses with frequency loci separated by a vast frequency gap. 
Perturbing such a two-color pulse compound triggers periodic amplitude and width variations, reminiscent of molecular vibrations.
With increasing strength of perturbation, the dynamics of the pulse compound changes from harmonic to nonlinear oscillations.
The periodic amplitude variations enable coupling of the pulse compound to dispersive waves, resulting in the resonant emission of multi-frequency radiation.
We demonstrate that the location of the resonances can be precisely predicted by phase-matching conditions.
If the pulse compound consists of a pair of identical subpulses, inherent symmetries lead to degeneracies in the resonance spectrum.
Weak perturbations lift existing degeneracies and cause a splitting of the resonance lines into multiple lines. 
Strong perturbations result in more complex emission spectra, characterized by well separated spectral bands caused by resonant Cherenkov radiation and additional four-wave mixing processes. 
\end{abstract}

%
%
%
%
%

\section{Introduction \label{sec:intro}}
Emission of radiation by physical objects is omnipresent in physical systems.
For instance, in classical electrodynamics, 
oscillating and linear accelerating point charges generate electric and magnetic fields.
Charged particles confined to circular orbits emit synchrotron radiation,  decelerating charged particles produce Bremsstrahlung, and
oscillating electric dipoles cause dipole-radiation characteristic of radiative transitions in atoms in molecules \cite{Smirnov:BOOK:2019}.

In the field of nonlinear optics many processes can be described by wave equations of the type of the nonlinear Schrödinger equation (NSE) \cite{Mitschke:BOOK:2016,Agrawal:BOOK:2019}, which, in the integrable case, supports localized field pulses given by solitary waves \cite{Drazin:BOOK:1989}. 
Such localized field pulses resemble particles, and can, under the right circumstances, sustain periodic variations of their amplitude and width.
The emission of resonance radiation caused by such oscillations has been studied in different nonlinear-optics settings.
%
For example, considering a NSE in which a varying dispersion coefficient perturbs transmitted pulses periodically,  dispersion-managed solitons where found to radiate through a resonance mechanism. The underlying parametric resonance and the radiative decay of dispersion-managed solitons were studied for rapidly varying dispersion maps \cite{Yang:PD:2001}, and for weakly varying dispersion maps \cite{Pelinovsky:JAM:2004}.
In dispersion oscillating fibers, such parametric resonances where investigated numerically and also experimentally for solitons (in case of anomalous dispersion) and shock fronts (in case of normal dispersion) \cite{Conforti:SR:2015}, demonstrating the generation of resonant radiation at multiple resonant frequencies.
%
In a standard NSE with added third-order disperion, higher-order solitons where shown to support a quite similar radiation mechanism \cite{Driben:OE:2015}. Therein, the periodic variation of the solitons peak intensity enabled the coupling of the localized state to continuum modes on the opposing side of the zero-dispersion point. As a result, resonant radiation in form of multi-frequency comb-like spectral bands is emitted. The generation of such resonant radiation was shown to also proceed in presence of the Raman-effect and pulse self-steepening \cite{Driben:OE:2015}.
%
In the case of the Lugiato-Lefever equation with third-order dispersion, i.e.\ a variant of the NSE with added damping an driving, a Hopf-bifurcation can result in oscillating dissipative solitons that shed resonant radiation in a series of well resolved spectral peaks \cite{Melchert:SR:2020}.

In analogy to quantum mechanics, where molecules are formed as bound-states of multiple atoms, nonlinear optics supports various types of molecule-like bound states of multiple field pulses.
For instance, usual soliton molecules which consist of two identical subpulses separated by a fixed time-delay, can be realized in a dispersion managed NSE \cite{Stratmann:PRL:2005,Hause:PRA:2008}. 
Mutually bound solitons also arise in models of coupled NSEs \cite{Ueda:PRA:1990,Menyuk:JOSAB:1988,Afanasjev:PZ:1988,Trillo:OL:88,Afanasjev:JQE:1989,Afanasjev:OL:1989,Mesentsev:OL:1992,Akhmediev:Chaos:2000}.
%
Specifically, for twin-core fibers with higher order dispersion, bound solitons near the zero-dispersion point where studied and shown to sustain periodic center-of mass and amplitude variations with different oscillation periods, giving rise to resonant radiation with different frequencies \cite{Oreshnikov:PRA:2017}.
Further, dissipative optical soliton molecules can be generated in  mode-locked fiber lasers \cite{Krupa:PRL:2017,Wang:NC:2019}.
%
Quite recently, a different kind of molecule-like pulse compound has been demonstrated theoretically \cite{Melchert:PRL:2019,Tam:PRA:2020}.
It consists of two subpulses at distinctly different center frequencies, which superimpose to form a single complex in the time-domain.  
An attractive, cross-phase modulation (XPM) induced potential holds these two-color pulses compounds together \cite{Melchert:PRL:2019}.
The binding mechanism that enables their stable propagation is different from that of usual soliton molecules. It requires a propagation constant with (at least) two separate domains of anomalous dispersion within which group-velocity matched co-propagation of pulses is possible.
A strong incoherent attractive interaction between the subpulses of such a two-color soliton, mediated accross (at least) two zero-dispersion points, is ensured if the group-velocity mismatch between the pulses is negligible \cite{Melchert:SR:2021}.
The emphasis on group-velocity matching is similar as for the repulsion between a soliton and a dispersive wave mediated across a single zero-dispersion point \cite{Philbin:S:2008,Demircan:PRL:2011,Faccio:CP:2012,Demircan:PRL:2013,Plansinis:PRL:2015}, where quasi group-velocity matching is vital to ensure a strong and efficient interaction mechanism.
In terms of a NSE with added negative fourth-order dispersion, two-color soliton molecules were identified as members of a large family of generalized dispersion Kerr solitons \cite{Tam:PRA:2020}.
Such two-color solitons were recently verified experimentally in mode-locked laser cavities \cite{Lourdesamy:NP:2021,Mao:NC:2021}. 
Two-color soliton microcomb states of similar structure also exist in the framework of the Lugiato-Lefever equation \cite{Melchert:OL:2020,Moille:OL:2018}.
The strong mutual interaction of the subpulses of such molecule-like bound states \cite{Willms:PRA:2022} renders them ideal for studying the effects of oscillations exhibited by coupled localized states in nonlinear optics systems.

Here, we discuss the rich dynamical behavior of such two-color soliton molecules, consisting of group-velocity matched pulses located in distinct domains of anomalous dispersion, separated by a vast frequency gap. 
We show that nonstationary dynamics of the subpulses results in emission
of resonant radiation.  The location of the newly generated multi-peaked
spectral bands can be understood by extending existing approaches
\cite{Yulin:OL:2004,Skryabin:PRE:2005,Conforti:SR:2015,Oreshnikov:PRA:2017,Melchert:SR:2020}
to two-frequency pulse compounds \cite{Oreshnikov:PP:2022}. Due to the manifold of two-frequency pulse
compounds with differing substructure, their emission spectra manifest in various complex forms.
Viewing a two-color soliton molecule as a superposition of two distinct
subpulses immediately suggests the existence of two different types of
dynamics: first, oscillations of the subpulses with respect to a common center
in time, and second, pure amplitude oscillations of the subpulses. 
In previous studies of two-color soliton molecules
\cite{Melchert:PRL:2019,Melchert:SR:2021,Willms:PRA:2022}, a combination of
both types of oscillations has generally been found. 
Therein, the first type of oscillation was observed to be heavily damped
\cite{Melchert:SR:2021,Willms:PRA:2022,Oreshnikov:PP:2022}. 
In the presented work we focus on pure amplitude oscillations, as they
represent the persistent oscillation mode of two-frequency soliton molecules.

The article is organized as follows.
In Sec.~\ref{sec:methods} we detail the propagation model used for our theoretical investigations and we specify initial conditions that directly generate special soliton-like two-color soliton pairs \cite{Melchert:OL:2021}, which we below also refer to as two-color soliton \emph{molecules} (or soliton molecules). This approach surpasses previous ``seeding'' procedures, which result in a largely uncontrolled generation of localized two-color pulse compounds.
In Sec.~\ref{sec:results} we will perturb such a two-color soliton molecule by systematically increasing its initial amplitude. In analogy to usual single-pulse nonlinear Schrödinger solitons, where increasing the amplitude results in an oscillating pulse, this triggers amplitude oscillations of the entire pulse compound, yielding a vibrating soliton molecule. We numerically demonstrate that the mutual driving of the subpulses results in the generation of resonant multi-frequency radiation and we compare the observed location of the resonances to analytic predictions obtained from a phase-matching analysis.
We demonstrate that symmetries, inherent in the propagation setting, can lead to degenerate resonance conditions, and we show how existing degeneracies can be lifted by a perturbation of the propagation constant.
Section~\ref{sec:conclusions} concludes with summary.

\section{Methods \label{sec:methods}}

\paragraph{Propagation model.}
For studying the propagation dynamics of oscillating two-color pulse compounds, we consider the modified nonlinear Schrödinger equation (NSE) 
\begin{eqnarray}
i \partial_z A = \left(\frac{\beta_2}{2} \partial_t^2 - \frac{\beta_4}{24}\partial_t^4\right) A - \gamma |A|^2 A, \label{eq:NSE}
\end{eqnarray}
for a complex-valued field $A\equiv A(z,t)$, on a periodic time-domain of extent $T$ with boundary conditions $A(z,-T/2)=A(z,T/2)$. In Eq.~(\ref{eq:NSE}), $\beta_2>0$ (in units of $\mathrm{fs^2/\mu m}$) is a positive-valued group-velocity dispersion coefficient, $\beta_4<0$ ($\mathrm{fs^4/\mu m}$) is a negative-valued fourth-order dispersion coefficient, and $\gamma$ ($\mathrm{W^{-1}/\mu m}$) is the nonlinear coefficient.
For the discrete set of angular frequency detunings $\Omega\in \frac{2\pi}{T}\mathbb{Z}$, the expressions
\begin{eqnarray}
&A_\Omega(z)=\mathsf{F}[A(z,t)] \equiv \frac{1}{T} \int_{-T/2}^{T/2} A(z,t)\,e^{i\Omega t}~{\rm{d}}t, \label{eq:FT_FT}\\
&A(z,t) = \mathsf{F}^{-1}[A_\Omega(z)] \equiv \sum_{\Omega} A_\Omega(z)\,e^{-i\Omega t}, \label{eq:FT_IFT}
\end{eqnarray}
specify forward [Eq.~(\ref{eq:FT_FT})], and inverse [Eq.~(\ref{eq:FT_IFT})] Fourier transforms that relate the field envelope $A(z,t)$ to the spectral envelope $A_\Omega(z)$.
Using Parseval's identity for these transforms, the total energy in both domains is given by
\begin{eqnarray}
E(z) = \int_{-T/2}^{T/2} |A(z,t)|^2~{\rm{d}}t = T \sum_\Omega |A_\Omega(z)|^2, \label{eq:E}
\end{eqnarray}
wherein $|A(z,t)|^2$ ($\mathrm{W=J/s}$) specifies the instantaneous power, and $|A_\Omega(z)|^2$ ($\mathrm{W}$) the power spectrum.

\begin{figure}[t!]
\includegraphics[width=\linewidth]{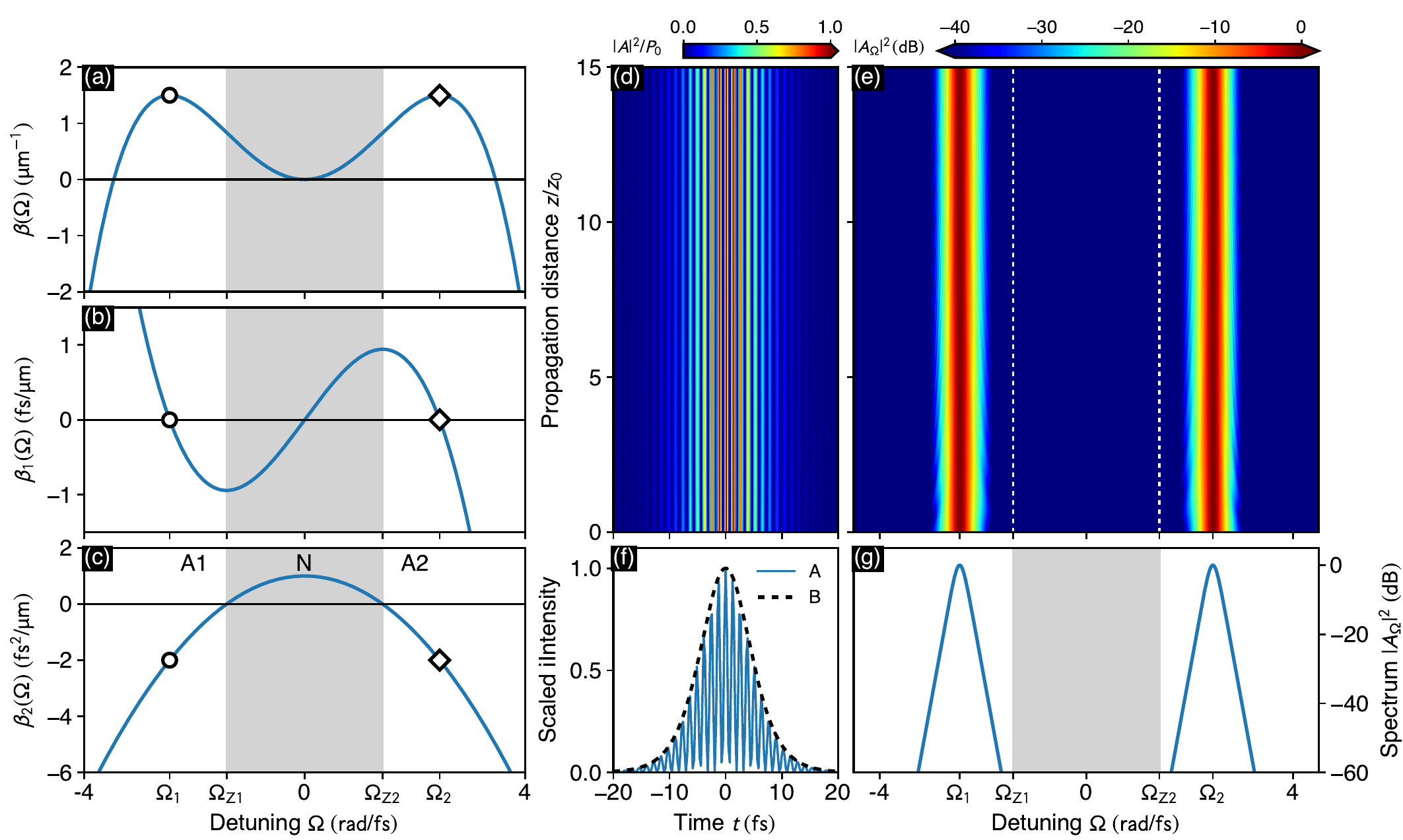}
\caption{Idealized setup yielding a two-color soliton molecule.
(a) Propagation constant, (b) inverse group velocity, and, (c) group-velocity dispersion. 
Circle and diamond highlight the two-color solitons subpulse loci. 
The domain of normal dispersion is shaded gray. In (c), A1 and A2 label distinct domains of anomalous dispersion, and N labels the domain of normal dispersion.  
(d) Intensity, and, (e) spectrum of a fundamental two-color soliton molecule for the propagation range $z=0\ldots 15 \, z_0$ with $z_0=t_0^2/(2 \beta_2)$. (f) Initial intensity, and, (g) initial spectrum at $z=0$. In (f), A labels the scaled intensity $|A|^2/P_0$, and B labels the squared magnitude of the meta-envelope $|F|^2/P_0={\rm{sech}}^2(t/t_0)$. 
\label{fig:01}}
\end{figure}

\paragraph{Propagation constant.}
Using the identity $\partial_t^n\,e^{-i \Omega t} = (-i\Omega)^n\,e^{-i\Omega t}$, 
the frequency-domain representation of the propagation constant is given by the polynomial expression $\beta(\Omega) = \frac{\beta_2}{2}\Omega^2 + \frac{\beta_4}{24} \Omega^4$. The inverse group-velocity of a mode at detuning $\Omega$ is given by $1/v_g(\Omega)=\beta_1(\Omega)\equiv \partial_\Omega \beta(\Omega)$, and the group-velocity dispersion is $\beta_2(\Omega)\equiv \partial_\Omega^2 \beta(\Omega)$.
Assuming $\beta_2>0$, and $\beta_4<0$, the group-velocity dispersion exhibits a downward parabolic
shape with zero-dispersion points defined by $\beta_2(\Omega)=0$ at $\Omega_{\rm{Z1},\rm{Z2}}=\mp
\sqrt{2\beta_2/|\beta_4|}$.
Subsequently we set $\beta_2 = 1~\mathrm{fs^2/\mu m}$, and $\beta_4= -1~\mathrm{fs^4/\mu m}$, resulting in the dispersion characteristics shown in Figs.~\ref{fig:01}(a-c). We further set $\gamma = 1~\mathrm{W^{-1}/\mu m}$.

\paragraph{Propagation algorithm.}
The numerical simulations reported in Sec.~\ref{sec:results} are performed using an algorithm with adaptive stepsize control, implementing the ``Conservation quantity error'' method (CQE) \cite{Heidt:JLT:2009,Melchert:CPC:2022}, which employs a
conservation law of the underlying propagation equation to guide selection of the propagation stepsize $h$. 
Specifically, we use the relative error 
\begin{eqnarray}
\delta_{E}(z) =  \frac{|E(z+h) - E(z)|}{E(z)}, \label{eq:dE}
\end{eqnarray} 
where $E$ is the energy~(\ref{eq:E}), conserved by the NSE~(\ref{eq:NSE}).
The CQE method keeps $\delta_{E}$ within the goal
error range $[0.1\,\delta_{\rm{G}}, \delta_{\rm{G}}]$, specified by the local goal error $\delta_{\rm{G}}$, by decreasing $h$ when necessary while
increasing $h$ when possible. For our numerical experiments we use
$\delta_{\rm{G}}=10^{-10}$.
To advance the field from position $z$ to $z+h$, we use the fourth-order
``Runge-Kutta in the interaction picture'' method (RK4IP) \cite{Hult:JLT:2007}.

\paragraph{Spectrograms.}
To assess the time-frequency interrelations within the field $A$
at a selected propagation distance $z$, we use the spectrogram
\cite{Melchert:SFX:2019,Cohen:IEEE:1989}
\begin{equation}
P_{S}(t,\Omega) = \frac{1}{2 \pi} \left|\int_{-T/2}^{T/2} A(z,t^\prime)h(t^\prime-t) e^{-i \Omega t^\prime}~{\rm d}t^\prime\right|^2. \label{eq:PS}
\end{equation}
Therein, $h(x)=\exp(-x^2/2\sigma^2)$ specifies a Gaussian window function with
root-mean-square width $\sigma$, used to localize the field in time.

\paragraph{Initial conditions.} 
Given the modified NSE~(\ref{eq:NSE}), two-color pulse compounds can be ``seeded'' by using initial conditions consisting of two superimposed fundamental solitons in the form
\begin{eqnarray}
A_0(t) = a_1 \,{\rm{sech}}(t/t_1)\, e^{-i \Omega_1 t} + a_2\, {\rm{sech}}(t/t_2)\,e^{-i\Omega_2 t}, \label{eq:IniCond} 
\end{eqnarray}
where $a_n^2 = |\beta_2(\Omega_n)|/(\gamma \,t_n^2)$, with $n\in(1,2)$, are the peak intensities of the two subpulses which are subject to the group-velocity matching condition $\beta_1(\Omega_1)=\beta_1(\Omega_2)$.
Upon propagation, such an initial condition will evolve into a narrow localized pulse which can exhibit internal dynamics, reminiscent of molecular vibrations, and is accompanied by free radiation \cite{Melchert:PRL:2019}. 
%
%
Recently we demonstrated that if the spectral extent of each subpulse is small compared to the frequency gap separating them, the mutual interaction of both pulses is effectively incoherent, allowing to construct special solutions in the form of two-color soliton pairs \cite{Melchert:OL:2021}. 
Considering the propagation constant used in Eq.~(\ref{eq:NSE}), an entirely symmetric two-color soliton pair can be constructed by choosing both subpulse loci at the two group-velocity matched frequencies $\Omega_{1,2}=\mp\sqrt{6\beta_2/|\beta_4|}$, highlighted in Figs.~\ref{fig:01}(a-c). They exhibit $\beta_1(\Omega_{1})=\beta_1(\Omega_2)=\beta_1(0)=0$ [Fig.~\ref{fig:01}(b)], and $\beta_2(\Omega_1)=\beta_2(\Omega_2)=-2 \beta_2 < 0$ [Fig.~\ref{fig:01}(c)]. Assuming two identical subpulse envelopes in Eq.~(\ref{eq:IniCond}), realized by $t_1=t_2\equiv t_0$ and $a_1^2=a_2^2=2\beta_2/(3 \gamma t_0^2)$ \cite{Melchert:OL:2021}, then yields 
\begin{equation}
    A_0(t)= F(0,t)\,\cos( \sqrt{2 \beta_2/|\beta_4|}t),\label{eq:TCSM}
\end{equation}
wherein 
\begin{equation}
F(z,t)= \sqrt{P_0}\,{\rm{sech}}(t/t_0)\,e^{i \kappa z} \label{eq:F}    
\end{equation}
specifies a pulse-enclosing envelope with peak power $P_0=8 \beta_2/(3 \gamma t_0^2)$ and nonlinear wavenumber $\kappa=\beta_2/t_0^2$  \cite{Melchert:OL:2021}.
These two-color soliton pairs correspond to the fundamental metasoliton describing generalized dispersion Kerr solitons, recently obtained in terms of a multi-scales analysis \cite{Tam:PRA:2020}.

\paragraph{Fundamental two-color soliton molecule.}
In Figs.~\ref{fig:01}(d-g), we demonstrate such a two-color soliton pair, having $\Omega_{1,2}=\mp \sqrt{6}\,\mathrm{rad/fs} \approx \mp 2.449~\mathrm{rad/fs}$, and $t_0=6~\mathrm{fs}$. Specifically, Fig.~\ref{fig:01}(f) shows the meta-envelope, enclosing the initial field. The vast frequency gap between the subpulse loci is evident in Fig.~\ref{fig:01}(g). 
The above condition for incoherent coupling, requiring the 
frequency separation of the pulses, given by $\Omega_{\rm{sep}}\equiv |\Omega_2-\Omega_1|= \sqrt{24 \beta_2/|\beta_4|}$, to be large compared to the
spectral extent of the individual subpulses, i.e.\ $\Omega_0 = 2/(\pi t_0)$, is clearly 
satified: the example in Fig.~\ref{fig:01} has $\Omega_{\rm{sep}}/\Omega_0 \approx 46$.
The soliton-like propagation of this fundamental two-color soliton pair is demonstrated in Figs.~\ref{fig:01}(d,e).

\section{Results \label{sec:results}}

To investigate the radiation that emanates from an oscillating two-color soliton molecule, we need to realize stable $z$-periodic dynamics. 
Specifically, in Sect.~\ref{ssec:res01} we demonstrate how to induce amplitude and width oscillations of soliton molecules and we study their characteristics.
In Sect.~\ref{ssec:res02} we demonstrate that oscillating soliton molecules shed resonant radiation in the form of multi-peaked spectral bands, and we compare our numerical results to predictions based on a phase matching analysis. 
In Sect.~\ref{ssec:res03} we show that possible degeneracies in the resulting radiation can be lifted by perturbations.

\subsection{Oscillating two-color soliton molecules \label{ssec:res01}}

\begin{figure}[t!]
\includegraphics[width=\linewidth]{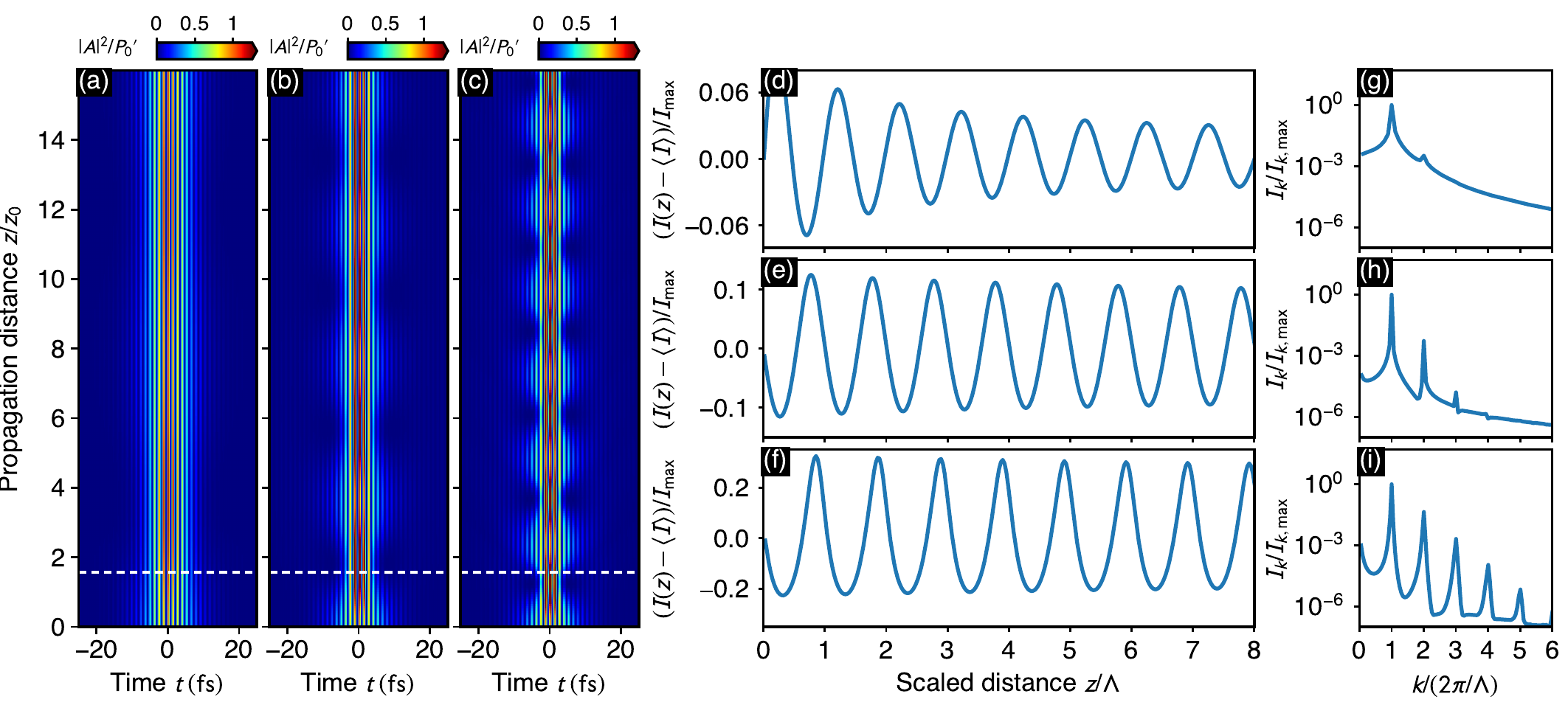}
\caption{
Vibration characteristics of two-color soliton molecules.
(a-c) Time-domain propagation dynamics for (a) $N=1.1$, (b) $N=1.4$, and, (c) $N=1.65$.
The horizontal dashed lines indicate $z=(\pi/2) \,z_0$, i.e.\ the oscillation period of a higher-order standard NSE soliton with dispersion length $z_0$.
(d-f) Variation of the scaled peak intensity followed for 8 consecutive oscillation periods for (d) $N=1.1$ with $\Lambda=4.30\,z_0$, (e) $N=1.4$ with $\Lambda=1.93\,z_0$, and, (f) $N=1.65$ with $\Lambda=1.13\,z_0$.
(g-i) Spatial Fourier modes corresponding to the intensity variations shown in (d-f).
\label{fig:02}}
\end{figure}

\paragraph{Amplitude oscillations.}
So as to excite amplitude and width oscillations of two-color soliton molecules, we consider initial conditions of the form of Eqs.~(\ref{eq:TCSM},\ref{eq:F}), and include an additional dimensionless factor $N$ through $P_0\to P_0^\prime \equiv N^2  P_0$, allowing to increase thir initial amplitude. 
The choice $N=1$ yields the basic soliton molecule of Eq.~(\ref{eq:TCSM}) [Fig.~\ref{fig:01}(f)], for which the pulse-enclosing envelope $F$ is coined the \emph{fundamental metasoliton} in Ref.~\cite{Tam:PRA:2020}.
%
In analogy with the nomenclature of usual single-pulse nonlinear Schrödinger solitons, choosing $N>1$ define higher-order metasolitons \cite{Tam:PRA:2020}, resulting in oscillating soliton molecules.
For brevity, we will refer to $N$ simply as the \emph{order} of the soliton molecule.
Figure~\ref{fig:02} demonstrates such oscillating soliton molecules for $N=1.1$ [Fig.~\ref{fig:02}(a)], $N=1.4$ [Fig.~\ref{fig:02}(b)], and $N=1.65$ [Fig.~\ref{fig:02}(c)].
If the initial perturbation of the two-color soliton molecule is small ($N=1.1$), the variation of the scaled peak intensity is largely harmonic [Fig.~\ref{fig:02}(d)], i.e.\ a single spatial Fourier mode describes the observed intensity variation [Fig.~\ref{fig:02}(g)].
As evident from Figs.~\ref{fig:02}(e,h) and \ref{fig:02}(g,i), increasing the molecule order results in nonlinear peak intensity oscillations, and, consequently, an increasing number of spatial Fourier modes.
Thereby, the variation of the molecule peak intensity is characterized by an oscillation period $\Lambda$ that decreases with increasing order $N$.

\begin{figure}[t!]
\centerline{\includegraphics[width=0.5\linewidth]{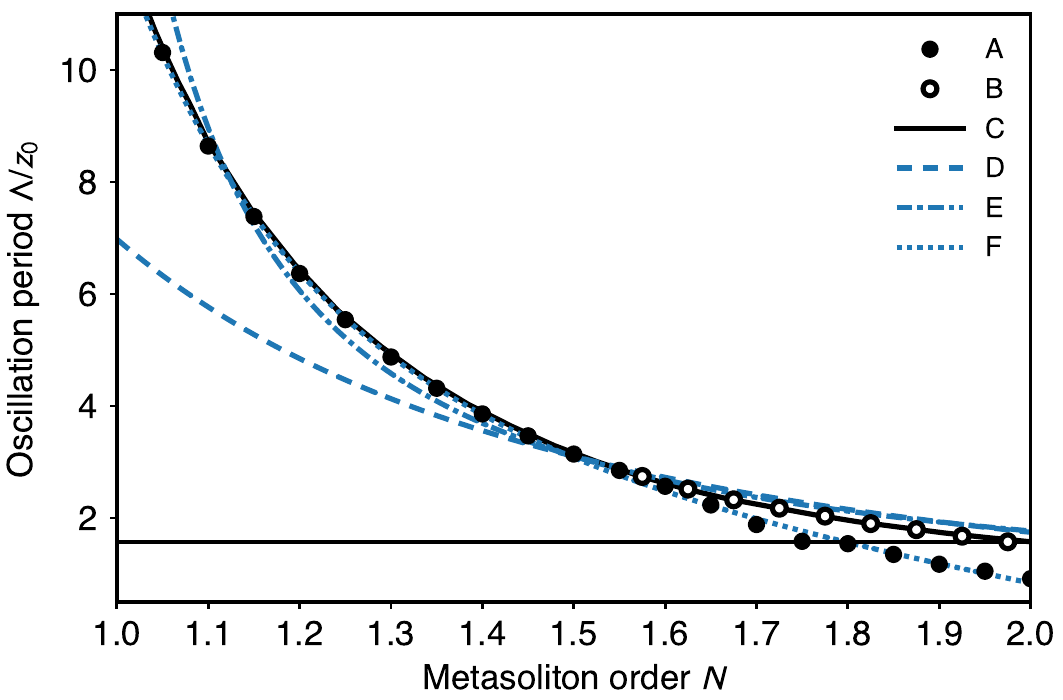}}
\caption{
Change of the oscillation wavelength $\Lambda$ with soliton molecule order $N$.
Data points, labeled A and B, show numerical results for soliton molecules with initial duration $t_0=6\,\mathrm{fs}$ and $t_0=12\,\mathrm{fs}$, respectively.
The oscillation period is scaled by $z_0=t_0^2/(2\beta_2)$.
Solid line, labeled C, shows numerical results for
standard NSE solitons of order $N$.
Dashed line, labeled D, shows results obtained by assuming a harmonic oscillation of the molecules subpulse widths.
Dash-dotted line, labeled E, shows the approximate oscillation period obtained from a variational approach.
Short dashed line, labeled F, 
shows a fit of Eq.~(\ref{eq:z_fit}) to the data for
molecules with $t_0=6\,\mathrm{fs}$.
\label{fig:03}}
\end{figure}

\paragraph{Amplitude oscillation wavelength.}
To quantify these dynamic amplitude and width variations we performed a series of numerical simulations for soliton molecules with different initial duration and order in the range $[1.05,2]$, see Fig.~\ref{fig:03}.
For $N \to 1$, the observed oscillation period agrees well with that of standard nonlinear Schrödinger solitons of order $N$ governed by a standard NSE, indicated by the solid line (labeled C) in Fig.~\ref{fig:03}.
This finding is reasonable, since, for $N \approx 1$, the coupling between the molecules subpulses remains effectively incoherent so that a given subpulse can be described by a standard NSE with modified coefficients \cite{Melchert:OL:2021}.
For $N\to 2$, the oscillation period of standard NSE solitons approaches the value $z_{\rm{sol}}= (\pi/2) z_0$, i.e.\ the soliton period of a higher-order (single) soliton with dispersion length $z_0$, indicated by the horizontal line in Fig.~\ref{fig:03}.
We find that the oscillation period of soliton molecules with large duration $t_0=12\,\mathrm{fs}$, labeled B in Fig.~\ref{fig:03}, are in excellent agreement with this behavior.
This finding is consistent with results obtained for a $N=2$ metasoliton of duration $t_0=\sqrt{2 |\beta_4|/(3 \beta_2 \epsilon^2)}\approx 16.3\,\mathrm{fs}$ ($\epsilon=0.05$), studied in Ref.~\cite{Tam:PRA:2020}.
In contrast, soliton molecules with smaller duration $t_0=6\,\mathrm{fs}$, labeled A in Fig.~\ref{fig:03}, deviate from this behavior for $N\gtrapprox 1.6$.
From our numerical experiments we found that for soliton molecules with small duration, even small amplitude oscillations suffice to cause a molecule to adiabatically change its shape, affecting its oscillation period and explaining the observed trend of the data.
Overall, the close correspondence between the oscillatory behavior of soliton molecules and standard NSE solitons allows to directly transfer  approximate descriptions of the evolution of near-fundamental standard NSE solitons, obtained in terms of variational approaches \cite{Anderson:PRA:1983,Kath:PRE:1995}, to the present case. 
A simple qualitative model of the observed subpulse oscillations can be derived on basis of a coupled system of variational equations, governing the evolution of the parameters of a hyperbolic-secant trial pulse \cite{Kath:PRE:1995}.  From this, an equation of motion can be obtained that captures the variation of the pulse width about some preferred, stationary value (see \ref{sec:A}). In lowest order approximation, it describes a harmonic oscillation. 
The corresponding wavelength
\begin{equation}
\Lambda^{\rm{HO}}(N) = \frac{\pi^2}{\sqrt{2} N^2} z_0,
\end{equation}
derived in \ref{sec:A}, is in qualitative agreement with our numerical results, see the dashes line (labeled D) in Fig.~\ref{fig:03}.
A less intuitive, but more accurate estimate of the oscillation period can be obtained by using Gaussian trial functions to approximate NSE solitons in a variational analysis of the standard NSE, yielding the approximate oscillation period \cite{Anderson:PRA:1983}
\begin{equation}
\Lambda^{\rm{var}}(N) =\frac{\pi \sqrt{2} N^2}{(\sqrt{2}N^2-1)^{3/2}} z_0, \label{eq:z_var}  
\end{equation}
which is in good agreement with our numerical results, as shown by
the dash-dotted line (labeled E) in Fig.~\ref{fig:03}.
%
A fit of the molecule data to a polynomial model of the form 
\begin{equation}
\Lambda^{\rm{fit}}(N) = z_0 \sum_{n=0}^3 c_n N^{-n},\label{eq:z_fit}
\end{equation}
yields the parameters $(c_0,c_1,c_2,c_3)=(-15.1, 64.3, -93.2, 56.2)$ for soliton molecules with initial duration $t_0=6\,\mathrm{fs}$,
shown as short dashed line (labeled E) in Fig.~\ref{fig:03}.
For molecules with $t_0=12\,\mathrm{fs}$, a fit yields $(c_0,c_1,c_2,c_3)=(-3.8, 24.7, -47, 39)$.

\subsection{Degenerated multi-frequency radiation \label{ssec:res02}}
\begin{figure}[t!]
\centerline{\includegraphics[width=\linewidth]{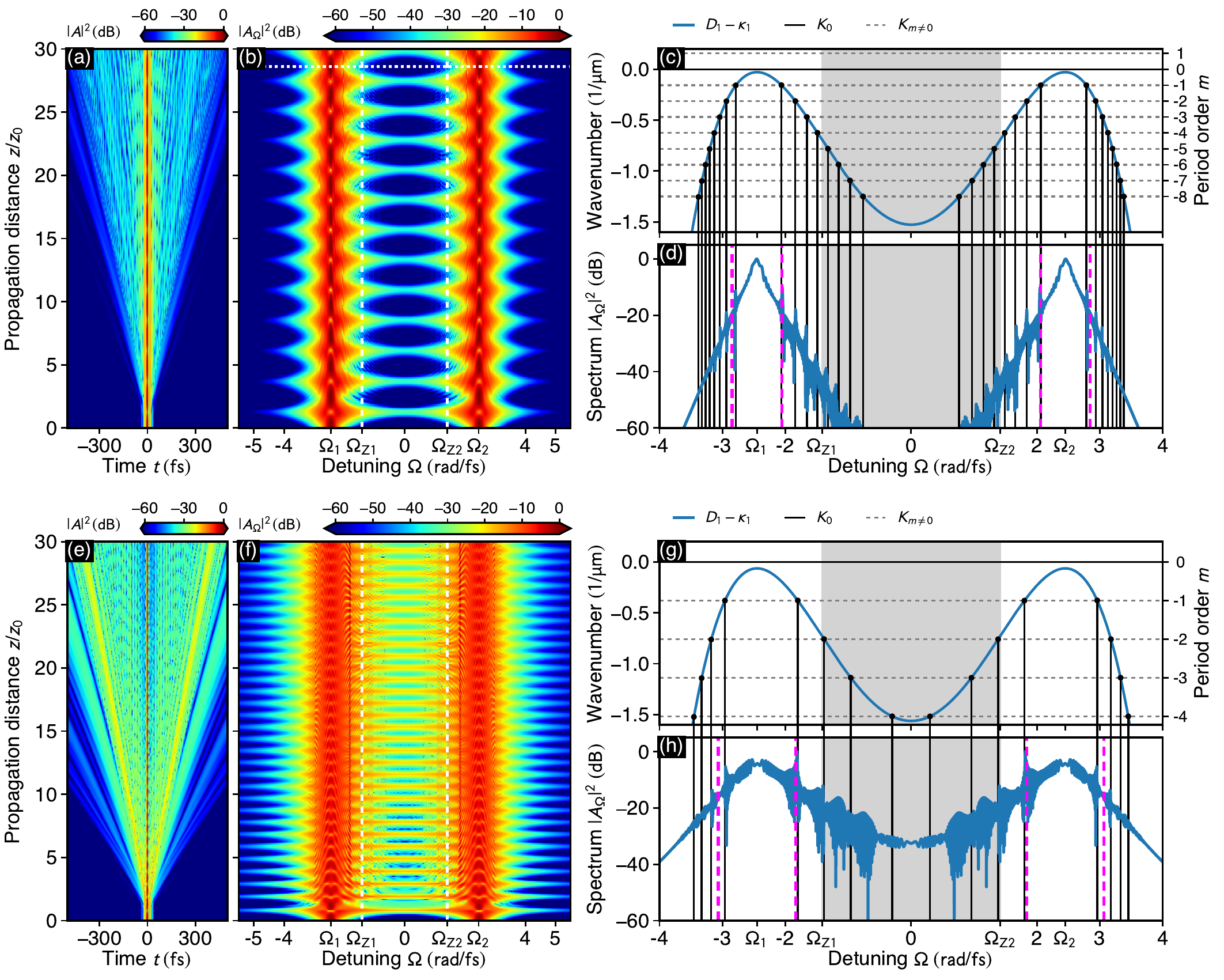}}
\caption{Resonant radiation of symmetric two-color soliton molecules.
(a-d) Results for soliton molecules of order $N=1.65$.
(a) Evolution of the intensity on a logarithmic scale. 
(b) Evolution of the spectrum. Horizontal short dashed line indicates $z/z_0=28.6$.
(c) Dispersion profile and graphical solution of the resonance conditions Eqs.~(\ref{eq:RR_01},\ref{eq:RR_02}) for $z$-oscillation periods of order $m$.
(d) Spectrum at $z/z_0\approx 28.6$.
Vertical lines in (c,d) indicate loci of resonant frequencies.
(e-h) Same for $N=2.0$. In (h) $z/z_0\approx 29.73$.
Vertical dashed lines in (d,h) indicate approximate loci of the resonances for $m=-1$, given by Eq.~(\ref{eq:Omega_RR_approx}).
\label{fig:04}}
\end{figure}

Below we show that the nonstationary dynamics of soliton molecules, which is reminiscent of molecular vibrations, results in emission of resonant multi-frequency radiation.
%
%
\paragraph{Example for $N=1.65$.}
As a first example, we consider a soliton molecule with duration $t_0=6\,\mathrm{fs}$, and order $N=1.65$. 
Figure~\ref{fig:04}(a) shows the time-domain intensity on a logarithmic scale [cf.~Fig.~\ref{fig:02}(c)], indicating that, in response to the initial perturbation, the localized molecule sheds radiation along either $t$-direction. 
The periodic variation of the spectrum, accompanying the amplitude and width oscillations of the molecule with wavelength $\Lambda/z_0=2.23$, is shown in Fig.~\ref{fig:04}(b). Distinct, comb-like bands of frequencies can be seen to develop in the vicinity of the subpulse loci $\Omega_1$ and $\Omega_2$.
These newly generated spectral bands appear particularly pronounced when the molecule assumes its maximum time-domain width during an oscillation period, so that the resonances are not drowned by the spectrum of the molecule, demonstrated for the specific choice $z/z_0 \approx 28.6$ in Figs.~\ref{fig:04}(b,d). 
The spectral location of this resonant radiation can be predicted in terms of wave number matching conditions, derived for the dynamically evolving, oscillating molecule state. For clarity, below we summarize the derivation of the resonance conditions detailed in Ref.~\cite{Oreshnikov:PP:2022}.
%
%
Considering an oscillating soliton molecule, consisting of two subpulses $U_1$ and $U_2$, we can account for amplitude and width oscillations by representing both in terms of a Fourier series 
 \cite{Oreshnikov:PP:2022}
\begin{equation}
U_n(z,t) = \sum_{m} C_{nm}(t)\,\exp\left(i\kappa_n z + im\frac{2\pi}{\Lambda}z\right), \quad n\in(1,2),~m\in\mathbb{Z} \label{eq:mol_vib_ansatz} 
\end{equation}
wherein $C_{nm}$ are expansions coefficients, $\kappa_n$ are the wavenumbers that govern the $z$-propagation of the subpulses, and $\Lambda$ is the wavelength of the periodic $z$-variation of the molecule. 
Below, we abbreviate the wavenumbers associated with the harmonics of the molecules $z$-oscillation wavelength $\Lambda$ by $K_m= 2\pi m/\Lambda$.
Under the assumption that these subpulses satisfy a pair of coupled NSEs \cite{Melchert:SR:2021,Melchert:OL:2021}, and by assessing how dispersive radiation is excited by their mutual periodic driving \cite{Oreshnikov:PP:2022}, the resonance conditions
\begin{eqnarray}
  &D_n(\Omega_{RR}) - \kappa_n = K_m, &\qquad n\in(1,2),~m\in\mathbb{Z} \label{eq:RR_01}\\
  &D_n(\Omega_{RR}) -2 \kappa_n + \kappa_{n^\prime} = K_m, &\qquad n, n^\prime \in (1,2),~n\neq n^\prime,~m\in\mathbb{Z} \label{eq:RR_02}
\end{eqnarray}
can be obtained.
Therein, $D_n(\Omega)\equiv \beta(\Omega)-\beta(\Omega_n) -\beta_1(\Omega)(\Omega-\Omega_n)$, for $n=1,2$, defines the dispersion profile in the reference frame of each of the subpulses, and $\Omega_{RR}$ specifies frequencies at which resonant radiation (RR) is excited.
The individual subpulse wavenumbers are related to the subpulse amplitudes $a_1$ and $a_2$ through
\begin{eqnarray}
\kappa_1 = \frac{\gamma}{2} (a_1^2 + 2 a_2^2),\label{eq:kap1}\\
\kappa_2 = \frac{\gamma}{2} (a_2^2 + 2 a_1^2) + \beta(\Omega_2)-\beta(\Omega_1)\label{eq:kap2}.
\end{eqnarray}
Equation~(\ref{eq:RR_01}) yields resonance conditions for the generation of Cherenkov radiation emitted by each of the molecules subpulses, related to the \mbox{$m$-th} harmonic of the driving period $\Lambda$.
Equation~(\ref{eq:RR_02}) defines further resonance conditions indicative of  four-wave mixing (FWM) processes that involve both subpulses.

In Ref.~\cite{Oreshnikov:PP:2022}, the subpulse wavenumbers $\kappa_1$ and $\kappa_2$
had to be extracted \emph{a posteriori} from the simulated propagation dynamics.
In the present case, we can establish a reasonable \emph{a priori} estimate by assuming the molecule to be close to a two-color soliton pair with subpulse amplitudes $a_1=a_2= \sqrt{2 \beta_2/ (3\gamma t_0^2)}$ \cite{Melchert:OL:2021} and $\beta(\Omega_1)=\beta(\Omega_2)=3\beta_2^2/(2 |\beta_4|)$, giving 
\begin{equation}
    \kappa_1 = \kappa_2 = \frac{\beta_2}{t_0^2}.
\end{equation}
Consequently, the resonance conditions (\ref{eq:RR_01}) and (\ref{eq:RR_02}) are degenerate.
The wavenumber profile in the reference frame of the subpulse at $\Omega_1$, given by $D_1(\Omega)-\kappa_1$, is shown in Fig.~\ref{fig:04}(c).  
In the reference frame of the subpulse at $\Omega_2$, the wavenumber profile $D_2(\Omega)-\kappa_2$ looks identical. This is a consequence of the symmetry of the propagation constant about the point $\Omega=0$, and the symmetric choice $-\Omega_1=\Omega_2=\sqrt{6}\,\mathrm{rad/fs}$ of the supulse loci.
A graphical solution of the resonance conditions~(\ref{eq:RR_01}) and (\ref{eq:RR_02}) can be obtained by superimposing the wavenumber profile with the wavenumbers $K_m$, associated with the harmonics of the molecules $z$-oscillation wavelength. 
This is demonstrated in Fig.~\ref{fig:04}(c), where $K_m$ for $m=-8\ldots 1$ are included as horizontal dashed lines. The crossing points, emphasized by dots in Fig.~\ref{fig:04}(c), indicate frequency loci $\Omega_{RR}$ at which the coupling of energy into dispersive waves is expected to be efficient.
As shown by the vertical lines in Figs.~\ref{fig:04}(c,d), we find that these resonant frequencies are in excellent agreement with our numerical results.
The large number of resonant frequencies, excited and continuously fed during propagation of the molecule, is related to the large number of modes required by a Fourier decomposition of the nonlinear vibration of the molecule for $N=1.65$ [cf.~Figs.~\ref{fig:02}(f,i)].
Molecules of order $N=1.4$ yield fewer and less pronounced resonances.
As shown by the spectrogram Fig.~\ref{fig:06}(a), all resonantly excited modes leave the molecule at a characteristic group-velocity, characteristic for their location in the vicinity of $\Omega_1$ and $\Omega_2$:
all resonantly excited modes in $[0,\Omega_1]$ and $[\Omega_2,\infty]$ propagate towards larger positive times; those in $[0,\Omega_2]$ and $[\Omega_1,\infty]$ propagate towards increasingly negative times.
We can simplify the above resonance condition to obtain a simple expression that specifies the location of the first few resonances reasonably well: by making a parabolic approximation of the propagation constant in the vicinity of $\Omega_1$ and $\Omega_2$, and by neglecting the contribution of the subpulse wavenumbers $\kappa_{1}$ and $\kappa_2$, we find 
\begin{equation}
\Omega_{\rm{RR},m}^{(1,2),\pm} \approx \Omega_{1,2} \pm \sqrt{\frac{|K_m|}{\beta_2}}.
\label{eq:Omega_RR_approx}
\end{equation}
The approximate positions of the resonances for $m=-1$ are indicated by vertical dashed lines in Figs.~\ref{fig:04}(d).

%
%
\paragraph{Example for $N=2$.}
According to Eq.~(\ref{eq:Omega_RR_approx}) it should be possible to control the separation between consecutive resonances by either decreasing $\beta_2$, resulting in a more ``flat'' dispersion profile $D_1(\Omega)$ in Fig.~\ref{fig:04}(c), or by increasing the wavenumbers $K_m$, resulting in more widely spaced horizontal dashed lines in Fig.~\ref{fig:04}(c).
To verify this prediction we next consider a soliton molecule of order $N=2$. Since such a molecule exhibits $\Lambda(N=2)<\Lambda(N=1.65)$ [Fig.~\ref{fig:03}], we expect to observe resonant frequencies with larger spacing as in the previously considered case.
As evident from \mbox{Figs.~\ref{fig:04}(e,f)}, the evolution of the molecule proceeds in two distinct stages: during the initial propagation stage $z/z_0 \lessapprox 10$, a large amount of dispersive radiation emanates from the molecule; then, for $z/z_0\gtrapprox 10$, the molecule seems to have reached a steadily oscillating stage.
The initial burst of energy results in an adiabatic reshaping of the molecule in course of which its temporal width decreases to about $t_0\approx 4\,\mathrm{fs}$, and its spectrum broadens. 
Consequently, the molecule also develops a notable spectral density within the domain of normal dispersion [Fig.~\ref{fig:04}(f)].
The wavelength of the $z$-periodic oscillations also decrease slightly from initially $\Lambda/z_0\approx 1.1$ to $\Lambda/z_0\approx 0.92$ for $z/z_0\gtrapprox 10$. 
This causes the wavenumbers $K_m$, indicated by horizontal dashed lines in Fig.~\ref{fig:04}(g), to shift downwards, and the corresponding resonant frequencies to shift away from the nearest subulse location. The associated spectral lines then appear broadened [Fig.~\ref{fig:04}(h)].
A similar observation has been made in a study of the resonance mechanism of higher-order solitons in a NSE with added third order dispersion \cite{Driben:OE:2015}: adding the Raman-effect, which causes a self-frequency-shift of solitons, was found to wash out the resonantly excited radiation bands. 
The approximate positions of the resonances for $m=-1$ are again indicated by vertical dashed lines in Figs.~\ref{fig:04}(h).
Contrasting the dynamics for $N=2$ with that for $N=1.65$ provides an unexpected observation: since the possible driving wavenumbers $K_m$ are now more widely spaced, see the horizontal dashed lines in Fig.~\ref{fig:04}(g), the molecule can support only fewer resonant frequencies; as a result, its amplitude and width oscillations appear more harmonic.
As evident from the spectrogram Fig.~\ref{fig:06}(b), due to the aforementioned decrease of $\Lambda$ for increasing $z$, some resonances, which are present during the initial ``nonlinearly oscillating'' propagation stage ($z/z_0\lessapprox 10$), die out and are absent in the later ``harmonically oscillating'' propagation stage ($z/z_0\gtrapprox 10$).
Let us note that for both cases considered above, the particular resonance for $m=0$, indicating the excitation of the usual Cherenkov radiation in absence of amplitude and width oscillations \cite{Akhmediev:PRA:1995}, is not excited.

In both cases studied above, we find that the amplitude and width oscillations of the soliton molecules are quasistationary and decay only slowly due to the emission of resonant radiation.

\begin{figure}[t!]
\centerline{\includegraphics[width=\linewidth]{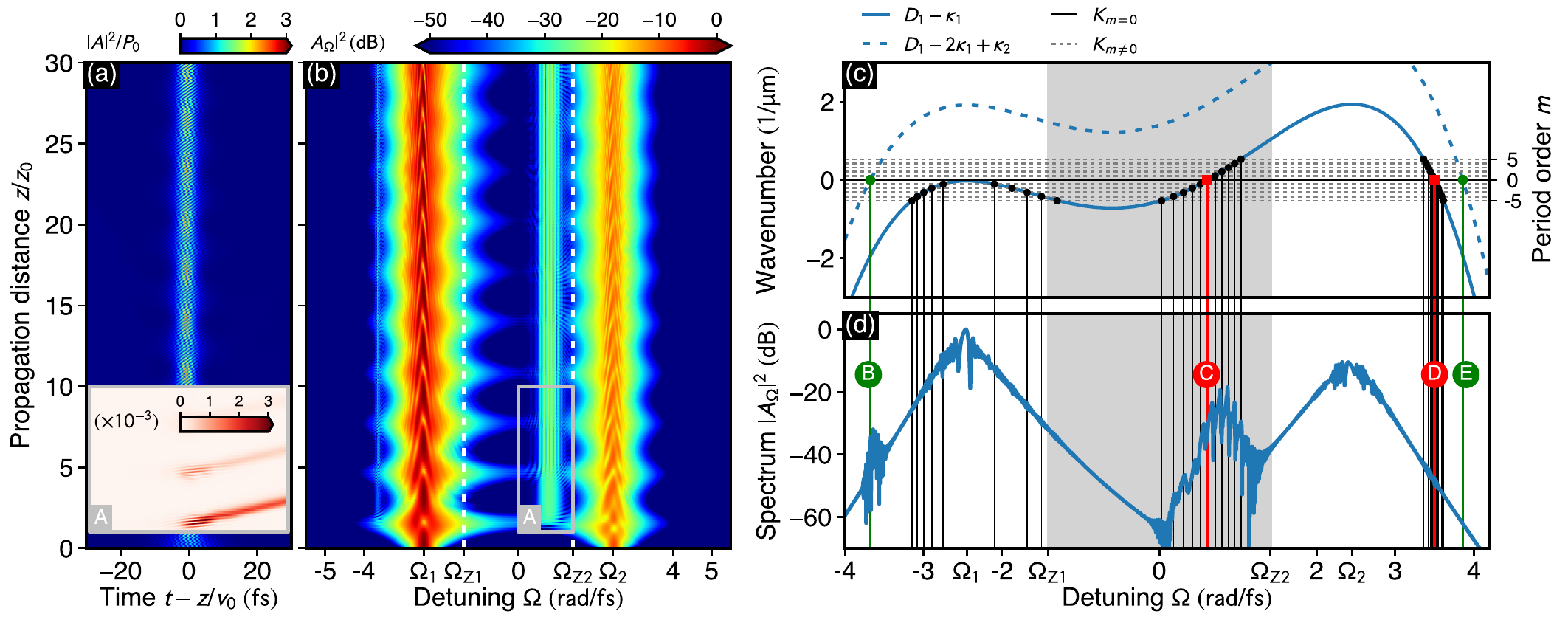}}
\caption{Resonant radiation of a two-color soliton molecule of order $N=1.65$ for a non-symmetric propagation constant, perturbed by $\beta_3=-0.2\,\mathrm{fs^3/\mu m}$.
(a) Evolution of the intensity on a linear scale. 
(b) Evolution of the spectrum.
A filtered view of the spectral components enclosed by the box (labeled A) in (b) is shown within the box (labeled A) in (a).
(c) Dispersion profile as seen for the subpulse at $\Omega_1$, and graphical solution of the resonance conditions~(\ref{eq:RR_01},\ref{eq:RR_02}) for $z$-oscillation periods of order $m=-5\ldots 5$. 
(d) Spectrum at $z/z_0\approx 28.5$.
Vertical lines in (c,d) indicate loci of resonant frequencies.
B and E label the loci of FWM resonances [Eqs.~(\ref{eq:RR_02})] expected in absence of oscillations. Similarly, C and D label those of Cherenkov resonances [Eqs.~(\ref{eq:RR_01})] in absence of oscillations. 
\label{fig:05}}
\end{figure}

\subsection{Lifting the degeneracies by perturbations \label{ssec:res03} }
In order to lift the degeneracy of the resonance conditions
(\ref{eq:RR_01},\ref{eq:RR_02}), which occurs for $\kappa_1=\kappa_2$,
Eqs.~(\ref{eq:kap1},\ref{eq:kap2}) suggest two possibilities:
either setting $a_1\neq a_2$ (see \ref{sec:B}), i.e.\ assigning different
amplitudes for both subpulses, 
or devising a propagation scenario for which $\beta(\Omega_1)\neq \beta(\Omega_2)$. 
Subsequently, we opt for the latter and assess the impact of a non-symmetric propagation constant on the dynamics of a soliton molecule.
Specifically, we modify the propagation constant to
$\beta(\Omega) = \frac{\beta_2}{2}\Omega^2 + \frac{\beta_3}{6} \Omega^3 + \frac{\beta_4}{24} \Omega^4$ using $\beta_3=-0.2\,\mathrm{fs^3/\mu m}$.
This breaks the symmetry inherent in the propagation scenarios studied above.
In Fig.~\ref{fig:05} we show how a soliton molecule with $N=1.65$, and subpulse center frequencies again at $\Omega_1=-\Omega_2=\sqrt{6}\,\mathrm{rad/fs}$,
responds to such a perturbation.
In this case, the effective values of second order dispersion felt by both subpulses are $\beta_2(\Omega_1)\approx-1.51\,\mathrm{fs^2/\mu m}$, and $\beta_2(\Omega_2)\approx -2.49\,\mathrm{fs^2/\mu m}$. 
We then start with a (fundamental) two-color soliton pair in the form of Eq.~(\ref{eq:IniCond}) with $a_1= \sqrt{(2\alpha-1)|\beta_2(\Omega_1)|/(3\gamma t_0^2)}$ and $a_2= \sqrt{(2\alpha^{-1}-1)|\beta_2(\Omega_2)|/(3\gamma t_0^2)}$, with $\alpha=\beta_2(\Omega_2)/\beta_2(\Omega_1)\approx 1.65$ \cite{Melchert:OL:2021}, and increase its amplitude by the factor $N$.
As a first observation, we note that the molecule acquires an additional shift in its group-velocity due to $\beta_3\neq 0$. In Fig.~\ref{fig:05}(a) we compensated for the resulting drift by transforming to a reference frame with velocity $v_0=-1.877\,\mathrm{\mu m/fs}$, within which the molecule appears stationary.
Further, the spectrum in Figs.~\ref{fig:05}(b,d) is now very different from  that in Figs.~\ref{fig:04}(b,d).
The soliton molecule exhibits a non-symmetric spectrum, consisting of a strong pulse, centered at $\Omega_1$, and a weaker pulse, centered at $\Omega_2$.
In such a scenario, due to a strong entanglement of the subpulses \cite{Willms:PRA:2022}, we expect that the strong pulse drives the weak pulse so that the wavelength of the intensity and width oscillations are equal for both pulses. From our numerical simulation we find $\Lambda/z_0 = 3.2$.
Below we perform a phase-matching analysis in the frame of reference of the stronger subpulse (i.e.\ the one at $\Omega_1$).
In this case, we have to resort to measuring the subpulse wavenumbers $\kappa_1$ and $\kappa_2$ \emph{a posteriori} from the simulated dynamics.
A graphical solution of the resonance conditions~(\ref{eq:RR_01}) for $m=-5\ldots 5$ shows that
now there are also multi-peaked spectral bands of resonances excited around $m=0$, indicated by the red short-dashed line in Fig.~\ref{fig:05}(c).
A filtered view of the resonant radiation at $\Omega\approx 0.6\,\mathrm{rad/fs} $, enclosed by a box (labeled A) in Fig.~\ref{fig:05}(b), is shown in the box (labeled A) in Fig.~\ref{fig:05}(a). As evident from this filtered view, the resonant radiation is emitted in a pulse-wise fashion and with decreasing intensity for increasing propagation distance \cite{Cristiani:OE:2004}. The phase-relationship between theses separated pulses then yields the multi-peaked spectral band in Fig.~\ref{fig:05}(d). 
Additional bands of resonances, associated with Cherenkov resonances, are visible at $\Omega\approx \Omega_1$ and $\Omega\approx 3.5\,\mathrm{rad/fs}$. 
For clarity, we provide a graphical solution of the phase-matching conditions~(\ref{eq:RR_01}) for $m=-5\ldots 5$, only.
Similar effects for perturbation by $\beta_3$ have previously been reported for higher-order solitons in the NSE \cite{Driben:OE:2015}, 
and dissipative solitons in fiber cavities \cite{Melchert:SR:2020}. 
Additional four-wave mixing (FWM) resonances arising from Eq.~(\ref{eq:RR_02}) can be identified at 
$\Omega\approx -3.8\,\mathrm{rad/fs}$. In contrast to an earlier approach, where such resonances where caused by the nonlinear mixing of solitons and dispersive waves \cite{Skryabin:PRE:2005}, we here find that these resonances are caused by a mixing of the subpulses of a soliton molecule.

\begin{figure}[t!]
\centerline{\includegraphics[width=\linewidth]{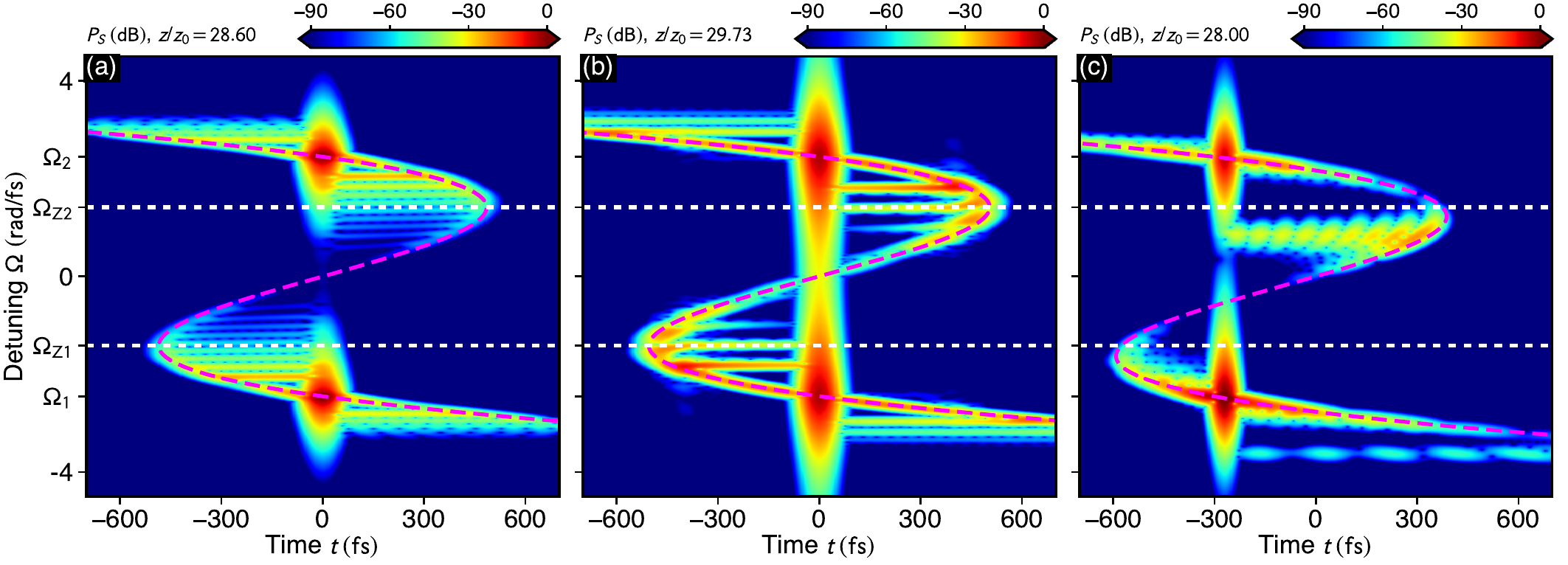}}
\caption{Spectrograms of vibrating soliton molecules and their resonantly shed radiation.
(a) Spectrogram of the oscillating molecule for $N=1.65$
at $z/z_0\approx 28.6$ [Figs.~\ref{fig:04}(a-d)], showing many narrowly spaced resonances. The spectrogram is computed for $\sigma=24\,\mathrm{fs}$ in Eq.~(\ref{eq:PS}).
(b) Spectrogram ($\sigma=24\,\mathrm{fs}$) of a molecule for $N=2$ at $z_0/z=29.73$ [Figs.~\ref{fig:04}(e-h)], showing few widely spaced resonances.
(c) Spectrogram ($\sigma=16\,\mathrm{fs}$) for a molecules with $N=1.65$ and a non-symmetric propagation constant [Fig.~\ref{fig:05}].
% ... 
The horizontal short-dashed lines indicate zero-dispersion points.
The dashed line indicates $\beta_1(\Omega)\,z$, i.e.\ the light-cone delimiting temporal position of a mode at $\Omega$ emitted at $z=0$.
Movies of the dynamics are provided as Supplementary Material.
\label{fig:06}}
\end{figure}

All of the resonantly generated radiation emanates from the molecule in the vicinity of the Cherenkov, and FWM resonances specified by Eqs.~(\ref{eq:RR_01},\ref{eq:RR_02}). The pulse-wise emission is clearly evident in the spectrogram Fig.~\ref{fig:06}(c). In Fig.~\ref{fig:06}(c) we did not compensate for the non-vanishing group velocity of the molecule, so that, at $z/z_0\approx 28$, it has drifted to $t\approx -300\,\mathrm{fs}$.
In the non-oscillating case, the observed multi-frequency spectral bands would degenerate into single, sharp spectral lines \cite{Akhmediev:PRA:1995,Yulin:OL:2004,Skryabin:PRE:2005}. Specifically, the points in the spectrum labeled C and D in Fig.~\ref{fig:04}(d) correspond to the location of Cherenkov radiation, and the points labeled B and E correspond to additional FWM processes in absence of oscillations.
Finally, let us note that in the present propagation scenario we do not observed resonant radiation in the vicinity of the point labeled E in Fig.~\ref{fig:05}(d). As discussed in Refs.~\cite{Skryabin:PRE:2005,Oreshnikov:PP:2022}, this can be attributed to a low efficiency of the corresponding resonant process. 
In comparison to the earlier two cases studies in Sect.~\ref{ssec:res02}, we here find that the amount of resonant radiation emanating from the soliton molecule is larger and that the amplitude oscillations decay faster.
Let us note that lifting the degeneracies between the resonance conditions
(\ref{eq:RR_01},\ref{eq:RR_02}) through weak perturbations first leads to a
split-up of the resonance lines into multiple lines (see \ref{sec:B}).  Only
when increasing the strength of the perturbation, complex spectra as in
Fig.~\ref{fig:05} are obtained.

As pointed out above, a viable means to lift existing degeneracies is to
consider pulses at a group velocity matched pair of frequencies for which
$\beta(\Omega_1)\neq \beta(\Omega_2)$. 
Above, this was achieved by considering a non-symmetric propagation constant.
Similarly, this can be achieved by retaining a symmetric propagation constant
but shifting the subpulses to a group-velocity matched pair of frequencies
different from the specific pair $\Omega_{1,2}$ considered above.
In addition, allowing a frequency dependence of the nonlinear coefficient
$\gamma$ in the nonlinear part of Eq.~(\ref{eq:NSE}) directly breaks the
symmetry of the subpulses that comprise a molecule. Propagation models of  this
type were considered in previous studies in Refs.\
\cite{Melchert:PRL:2019,Melchert:SR:2021,Willms:PRA:2022,Oreshnikov:PP:2022}.

\section{Summary and conclusions \label{sec:conclusions}}
We studied two-color soliton molecules, i.e.\ pulse compounds consisting of two tightly bound subpulses at vastly distinct frequencies, supported by nonlinear waveguides with two distinct domains of anomalous dispersion. 

First, we analyzed the amplitude and width oscillations of such molecules, induced by increasing the molecule order in a systematic way. The order of a molecule is defined similar to the usual order of solitons in the standard NSE.
We contrasted the periodic dynamics exhibited by two-color soliton molecules with that of single solitons in the standard NSE, finding that, in the limit of large molecule duration and despite their complex internal structure, their $z$-oscillation wavelength matches that of a standard NSE soliton of the same order. 
By elaborating on a well established variational analysis of the NSE we were led to expect that near-fundamental soliton molecules oscillate harmonically, whereas the amplitude oscillations quickly grow nonlinear with increasing molecule order. This simple and intuitive picture was found to be in good qualitative agreement with our numerical results.

Second, we investigated the emission spectrum of soliton molecules, excited by their periodic amplitude and width oscillations. As soon as a molecule oscillates, it sheds resonant radiation in the form of multi-frequency spectral bands, reminiscent of the shape of traditional Japanese \emph{Kushi} combs. The frequency spacing of resonantly excited modes is determined by the $z$-oscillation wavelength of the molecule.
We demonstrated that the location of the resonant frequencies can be reliably predicted by means of resonance conditions resulting from a phase-matching analysis. 
Specifically, there are two types of resonance conditions which separately determine the generation of Cherenkov radiation and further FWM processes. 
We showed that if the propagation scenario is entirely symmetric, these resonance conditions are redundant, so that the generated radiation is degenerate.
Breaking the symmetry of the propagation scenario, which we here accomplished by perturbing the propagation constant, the degeneracy can be lifted and distinct multi-frequency spectral bands uniquely associated with Cherenkov radiation, and further FWM processes, are obtained. 

The occurrence of such multi-frequency radiation, especially in the degenerate form, constitutes a fundamental phenomenon in nonlinear waveguides. 
The presented study clarifies the mechanism underlying the generation of resonant radiation of two-color pulse solitons, observed previously in terms of theoretical studies \cite{Melchert:PRL:2019}, and recently also in experiments on multi-wavelength mode-locked solitons in laser cavities \cite{Mao:NC:2021}.
Our results provide further insight into the puzzling propagation dynamics and radiative processes in presence of two-frequency soliton molecules.
Due to the manifold of two-frequency pulse compounds with differing substructure, their emission spectra manifest in various complex forms.
This demonstrates a peculiar nonlinear optics system for investigating  emission of resonant radiation.

\ack
OM, SW, IB, UM, and AD acknowledge financial support from Deutsche
Forschungsgemeinschaft (DFG) under Germany’s Excellence Strategy within the
Cluster of Excellence PhoenixD (Photonics, Optics, and Engineering—Innovation
Across Disciplines) (EXC 2122, Project No.\ 390833453).
AY acknowledges financial support from Priority 2030 Academic Leadership Program and goszadanie no.\ 2019-1246. 
IB also acknowledges support from DFG (Project No.\ BA4156/4-2).
UM also acknowledges support from DFG (Project No.\ MO 850-20/1).

\appendix 

\section{Simplified description of soliton molecule amplitude and width oscillations\label{sec:A}}

Considering the standard NSE in nondimensional form
\begin{equation}
i\partial_z A = -\frac{1}{2}\partial_t^2 A - |A^2| A, \label{eq:sNSE}
\end{equation}
a variational approach based on the trial function 
\begin{equation}
A(z,t)=\eta\, {\rm{sech}}\left(\frac{t}{w}\right) \exp\left(i\theta + ib \frac{t^2}{2w}\right), \label{eq:trialFun}
\end{equation}
yields the set of coupled ordinary differential equations 
\begin{eqnarray}
&\frac{d}{dz}(\eta^2 w) = 0,\label{eq:eta}\\
&\frac{d}{dz} w = b,\label{eq:ODE_w}\\
&\frac{d}{dz} b = \frac{4}{\pi^2 w^3} (1-\eta^2 w^2),\label{eq:ODE_b}\\
&\frac{d}{dz} \theta = \frac{5 \eta^2}{6} - \frac{1}{3 w^3},
\end{eqnarray}
describing the $z$-variation of the amplitude ($\eta$), pulse width ($w$), 
pulse phase ($\theta$), and, chirp ($b$)
of the trial function~(\ref{eq:trialFun}) \cite{Kath:PRE:1995}.
Equation~(\ref{eq:eta}) implies the constant of motion $\eta^2 w = \rm{const.}$, linking intensity and width variations of the trial function~(\ref{eq:trialFun}).
For an approximate analysis of the pulse width variation along $z$,
we consider the $z$-derivative of Eq.~(\ref{eq:ODE_w}), which, together with Eq.~(\ref{eq:ODE_b}), defines the nonlinear equation of motion
\begin{equation}
   \frac{d^2}{dz^2} w = \frac{4}{\pi^2 w^3} (1-\eta^2 w^2)\equiv f(w). \label{eq:HO_1}
\end{equation}
Assuming a pulse width $w(z)=w_0 + \epsilon(z)$ that performs periodic oscillations with amplitude $\epsilon(z)$ about some preferred value $w_0$, we may expand Eq.~(\ref{eq:HO_1}) up to first order in $\epsilon$ to find 
$\frac{d^2}{dz^2} \epsilon \approx f(w_0) + f^\prime(w_0)\, \epsilon$ with $f(w_0)= \alpha w_0^{-3} + \beta w_0^{-1}$, and $f^\prime(w_0)\equiv \frac{\partial}{\partial w} f(w)|_{w=w_0}=-3 \alpha w_0^{-4} - \beta w_0^{-2}$, with $\alpha =4 \pi^{-2}$ and $\beta=-\alpha \eta^2$.
Imposing the condition $f(w_0)=0$ fixes $w_0=\eta^{-1}$, i.e.\ a reasonable choice for $w_0$ given the trial function~(\ref{eq:trialFun}), and brings Eq.~(\ref{eq:HO_1})
to the form of a harmonic oscillator
\begin{eqnarray}
  \frac{d^2}{dz^2} \epsilon(z) + K^2(\eta)\, \epsilon(z) \approx 0,\label{eq:HO_2}
\end{eqnarray}
with angular wavenumber $K(\eta)=2\sqrt{2}\eta^2/\pi$. 
The wavelength of this harmonic oscillation then reads 
\begin{equation}
\Lambda^{\rm{HO}}(\eta) = \frac{2 \pi}{ K(\eta)}= \frac{\pi^2}{\sqrt{2} \eta^2}.\label{eq:zp_var_2} 
\end{equation}

We can readily transfer this result to the case of amplitude and width oscillations of two-color soliton molecules.
In the limit where the spectral extent of a molecules subpulses is small in comparison to the frequency gap separating them, i.e.\ under the condition $\pi t_0 \sqrt{6 \beta_2/|\beta_4|} \gg 1$,
each subpulse can effectively be modeled in terms of a standard NSE \cite{Melchert:OL:2021}. Then we are led to expect that the oscillation period of each soliton molecule subpulse, and thus the entire soliton molecule, is reasonably well described by the  approximate result Eq.~(\ref{eq:zp_var_2}). 
We plot $\Lambda^{\rm{HO}}(\eta)/z_0$, with $z_0=1$ for Eq.~(\ref{eq:sNSE}), as dashed line, labeled D, in Fig.~\ref{fig:03}.
Let us note that expanding Eq.~(\ref{eq:HO_1}) up to second order in $\epsilon$ yields a correction term $\propto [f^{\prime\prime}(w_0)/2]\epsilon^2$ with $f^{\prime\prime}(w_0)\equiv \frac{\partial^2}{\partial w^2}f(w)|_{w=w_0}=40\eta^5/\pi^2$, indicating that the anharmonicity of the oscillation increases rapidly with increasing amplitude. This finding is in qualitative agreement with the numerical results in Sec.~\ref{ssec:res01}.

\begin{figure}[t!]
\centerline{\includegraphics[width=\linewidth]{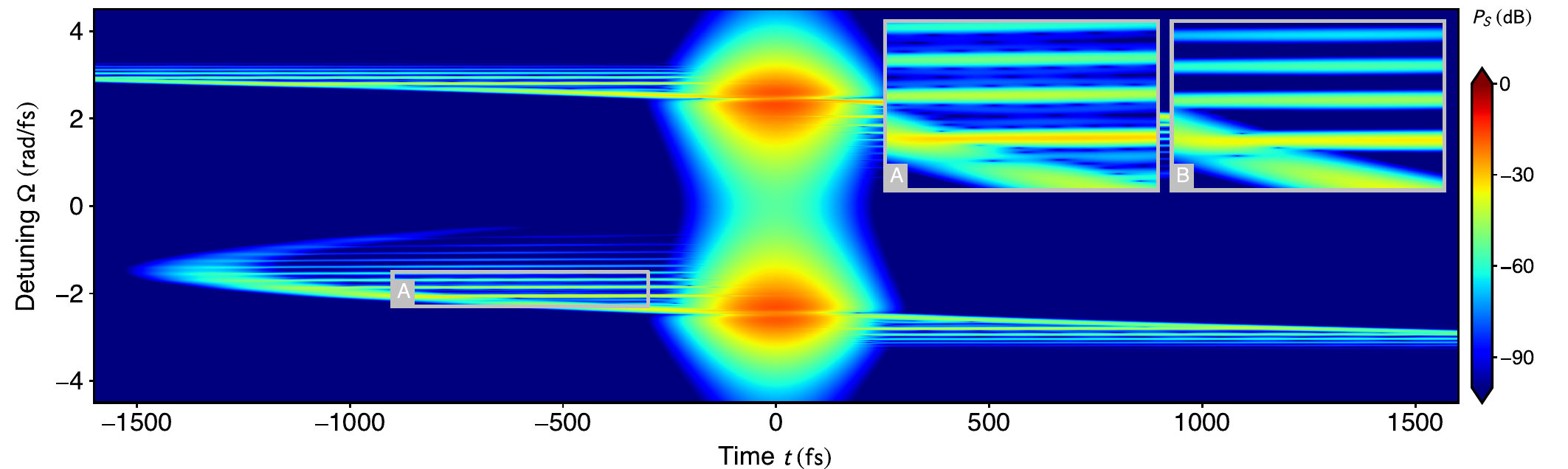}}
\caption{Spectrogram ($\sigma=70~\mathrm{fs}$) of a vibrating soliton molecule
with slightly different subpulse amplitudes.
The main plot shows the propagation dynamics starting from an initial condition
given by Eq.~(\ref{eq:IniCond_2}) with $N_1=1.59$ and $N_2=1.66$. Parts of the
main plot enclosed by the box is detailed in the close-up view labeled A.
Due to the unequal subpulse amplitudes, the degeneracy among the resonances 
conditions (\ref{eq:RR_01},\ref{eq:RR_02}) is lifted.
For comparison, the close-up view labeled B shows the corresponding plotting
range for the degenerate case with $N_1=N_2=1.59$.
\label{fig:07}}
\end{figure}

\section{Splitting of resonance lines in presence of small amplitude differences\label{sec:B}}

To complement the strong symmetry breaking perturbations induced by a
non-symmetric propagation constant, discussed in Sec.\ \ref{ssec:res03}, we
here demonstrate degeneracy-lifted multifrequency radiation resulting from a
small difference in the molecules subpulse amplitudes.
Specifically, we consider the symmetric propagation constant
$\beta(\Omega)=\frac{\beta_2}{2}\Omega^2 + \frac{\beta_4}{24}\Omega^4$ for
$\beta_2=1\,\mathrm{fs^2/\mu m}$ and $\beta_4=-1\,\mathrm{fs^4/\mu m}$, and 
the initial condition 
\begin{eqnarray}
A_0(t) = a_1 \,{\rm{sech}}(t/t_1)\, e^{-i \Omega_1 t} + a_2\, {\rm{sech}}(t/t_2)\,e^{-i\Omega_2 t}, \label{eq:IniCond_2} 
\end{eqnarray}
where $\Omega_{1,2}=\mp \sqrt{6}\,\mathrm{rad/fs }$, $t_0=6\,\mathrm{fs}$, and
$a_{1,2}^2 = N_{1,2}^2 2 \beta_2/(\gamma \,t_0^2)$, with $N_1=1.57$ and
$N_2=1.66$. 
Correspondingly, $a_2\gtrapprox a_1$ in Eqs.~(\ref{eq:kap1},\ref{eq:kap2}).
A spectrogram of the resulting vibrating soliton molecule at $z/z_0 \approx 80$
[$z_0=t_0^2/(2\beta_2)$] is shown in Fig.~\ref{fig:07}.
As evident from the spectrogram range $[\Omega=-2.3 \ldots -1.5~\mathrm{rad/fs}] \times
[t=-900\ldots -300~\mathrm{fs}]$, highlighted in close-up view A, the degeneracy of the
resonances conditions (\ref{eq:RR_01},\ref{eq:RR_02}) is lifted.
The primary resonances, resulting from solutions to Eq.~(\ref{eq:RR_01}), 
are now accompanied by pairs of secondary resonances, resulting from solutions 
to Eq.~(\ref{eq:RR_02}) for $n=1,2$, which appear on either side of a 
primary resonance.
To facilitate comparison to the degenerate scenario, close-up view B shows
the corresponding numerical results for $N_1=N_2=1.59$.

\section*{References}
\bibliographystyle{iopart-num}
\bibliography{references}

\end{document}